\documentclass[runningheads]{llncs}
\usepackage[T1]{fontenc}
\usepackage{graphicx}
\usepackage{listings}
\usepackage{comment}

\usepackage{array}
\newcolumntype{P}[1]{>{\centering\arraybackslash}p{#1}}

\begin{document}
\title{Designing a User Contextual Profile Ontology: A Focus on the Vehicle Sales Domain}
%
\titlerunning{Designing a User Contextual Profile Ontology}
%

\author{LE Ngoc Luyen\inst{1,2}, Marie-Hélène ABEL\inst{1}, Philippe GOUSPILLOU\inst{2}
}
\institute{Université de technologie de Compiègne, CNRS, Heudiasyc (Heuristics and Diagnosis of Complex Systems), CS 60319 - 60203 Compiègne Cedex, France \and Vivocaz, 8 B Rue de la Gare, 02200, Mercin-et-Vaux, France}
%
%
%
\maketitle              
\begin{abstract}
In the digital age, it is crucial to understand and tailor experiences for users interacting with systems and applications. This requires the creation of user contextual profiles that combine user profiles with contextual information. However, there is a lack of research on the integration of contextual information with different user profiles. This study aims to address this gap by designing a user contextual profile ontology that considers both user profiles and contextual information on each profile. Specifically, we present a design and development of the user contextual profile ontology with a focus on the vehicle sales domain. Our designed ontology serves as a structural foundation for standardizing the representation of user profiles and contextual information, enhancing the system's ability to capture user preferences and contextual information of the user accurately. Moreover, we illustrate a case study using the User Contextual Profile Ontology in generating personalized recommendations for vehicle sales domain.

\end{abstract}

\section{Introduction}\label{sec01introduction}

Understanding and customizing experiences for users interacting with various systems and applications is essential in today's digital landscape. To accomplish this, individual user profiles are integrated with their corresponding contextual information, resulting in the creation of a user contextual profile \cite{li2019using}. A user profile is a compilation of personal information and preferences about a particular user, including demographic data, interests, past behavior, and other attributes that define the user's identity within a system or application \cite{maria2007creating}. For example, the user Henri wants to buy a vehicle for professional use and another for his family with two children. Therefore, the user Henri can have two distinct profiles, each catering to different needs and preferences. While contextual information for each profile refers to the situational factors surrounding a user's interaction with a system or application \cite{tamine2010evaluation}. This may include location, time, device being used, and even the user's current activity or emotional state. In the example, each of the two profiles of the user Henri has unique contextual information related to Henri's preferences, needs, and interactions within the application. Consequently, a user contextual profile combines the user profile with the contextual information specific to that profile, creating a more comprehensive understanding of the user's needs and preferences in various situations. For instance, when searching for a professional vehicle, the user contextual profile may prioritize factors such as fuel efficiency, compact size, and reliability, while for a family vehicle, it may focus on safety features, ample seating, and cargo space.

In order to effectively model user contextual profiles, utilizing ontology-based approaches can be advantageous. Ontologies are formal, structured representations of knowledge within a specific domain that enable the organization and sharing of data in a machine-readable format. Once the ontology is designed and implemented, it can be integrated into applications and systems, allowing them to reason over the user contextual profiles and infer new knowledge based on the available data \cite{wang2004ontology,sutterer2008user,luyen2016development}. This can improve the accuracy and relevance of recommendations, as well as facilitate more efficient data integration and interoperability among different applications and services.

A considerable number of studies in the field of user modeling have primarily focused on either user profiles or user contexts \cite{sutterer2008upos,skillen2012ontological}. The aspects of contextual information related to each profile of the user have not been thoroughly examined.  This underexplored area offers an opportunity for further research, focusing on the integration of contextual information with different profiles of the user to develop a deeper understanding of users' needs and preferences in various situations. Consequently, our work aims to address this gap by designing a User Contextual Profile Ontology (UCPO) with a focus on the vehicle sales domain. This ontology considers both user profiles and contextual information on each profile, utilizing a standard ontology development methodology. Furthermore, we provide an illustration of the use of UCPO through a case study on generating personalized recommendations in the vehicle sales domain.

The remainder of this article is organized as follows: Section \ref{sec02relatedworks} introduces works from the literature on which our approach is based. Section \ref{sec03ourapproach} presents our main contributions, outlining the methodology for designing a user contextual profile ontology with a focus on the vehicle sales domain. In section \ref{sec04experiments}, we provide a case study on personalized recommendations based on our developed ontology. Finally, we conclude and discuss the perspectives.

\section{Related work}\label{sec02relatedworks}
In this section, we investigate the modeling and integration of user profiles and contextual information for specific profiles, as well as examine various approaches to ontology design. The related works' analysis will inform and support the development of a user contextual profile ontology.
\subsection{Ontological user modeling}

Ontologies have proven to be highly effective in modeling user profiles and contexts. They provide a thorough depiction of a particular domain of interest, simplifying browsing and query refinement. Specifically, ontologies have been observed to outperform other methods in user modeling when compared to other methods utilized \cite{sosnovsky2010ontological}. Ontological user modeling refers to the use of ontologies for representing and managing user profiles, preferences, and contextual information in a structured and machine-readable format. User-related information such as profiles, preferences, and context are represented using concepts, relationships, and attributes within a formal, hierarchical structure. This structured representation enables more effective processing, reasoning, and inferencing by computer systems.

Ontological user modeling has previously been investigated across numerous research areas, demonstrating its adaptability and potential to enhance user experiences in diverse domains such as e-learning \cite{kritikou2008user}, e-commerce \cite{he2008ontological}, recommender systems \cite{le2022apport,luyen2023towards,le2023personalized,luyen2023}, personalized web services \cite{sieg2007web}, and context-aware applications \cite{rimitha2019improving}, among others \cite{munoz2021gente,kourtiche2020oupip,le2023constraint}. Specifically, Golemati et al. \cite{maria2007creating} designed an ontology that incorporates concepts and properties essential for modeling user profiles. Their model predominantly focuses on static user characteristics and provides a foundation for the development of a more general, comprehensive, and extensible user model. 
Adewoyin et al. \cite{adewoyin2022user} proposed a versatile user modeling architecture for smart environments. They represented the characteristics of users using the behavior concept, which encompasses all relevant aspects that can aid in effective behavioral modeling and monitoring. Sutterer et al. \cite{sutterer2008upos} put forth a user profile ontology designed specifically to depict situation-dependent sub-profiles. This ontology can be utilized by context-aware adaptive service platforms for mobile communication and information services, facilitating automatic and situation-dependent personalization of such services.
 Skillen et al. \cite{skillen2012ontological} developed a User Profile Ontology for personalizing context-aware applications in mobile environments. Their focus lies on user behavior and characterizing the needs of users for context-aware applications. Contrasting with the work in \cite{maria2007creating}, their ontological user modeling approach emphasizes dynamic components for application usage. 
 
 The works mentioned highlight numerous advantages of ontological user modeling, which is capable of capturing both static and dynamic user characteristics related to permanent, temporary information, and user evolution. Moreover, user contexts play a crucial role in user modeling, contributing significantly to the creation of personalized and adaptive systems. Most of these studies integrate user contexts and user profiles. Nonetheless, the contextual information can be dependent on the profile and its primary objectives. Therefore, it is necessary to distinguish contextual information for the user from that for the profile clearly. The development of a user contextual profile ontology is essential for clearly organizing and distinguishing the contextual information related to users and their respective profiles. In the following section, we will examine ontology development methodologies to achieve this objective.
\subsection{Ontology development methodology}
Ontology development methodologies provide a structured and systematic approach to the creation, maintenance, and evolution of ontologies. These methodologies are specifically designed to ensure the quality, coherence, and practicality of the resulting ontology, while also promoting its consistency and usability. By employing ontology development methodologies, researchers and practitioners can effectively enhance the overall performance and applicability of the ontologies they create, making them more reliable and valuable for their intended use cases. 

Various ontology development methodologies have been proposed in the literature, each featuring its own unique set of guidelines and principles. Some widely adopted methodologies include: Methontology \cite{fernandez1997methontology}, Ontology Development 101 (OD101) \cite{noy2001ontology}, NeON Methodology \cite{suarez2011neon}, Unified Process for Ontology Building (UPON) \cite{de2005proposal}, Simplified Agile Methodology for Ontology Development (SAMOD) \cite{peroni2017simplified}, Modular Ontology Modeling (MoMo) \cite{shimizu2021modular}, and others. Firstly, Methontology is one of the earliest and most widely used methodologies. It covers the entire ontology development process, including specification, conceptualization, formalization, integration, implementation, and maintenance, while emphasizing the importance of documentation throughout the development process. OD101 then is a methodology designed to guide ontology engineering processes. Focusing on iterative development, OD101 provides step-by-step guidelines and best practices for constructing ontologies, ensuring a systematic and efficient approach to ontology creation. The NeON Methodology next is designed to create ontology networks, rather than singular ontologies. This approach emphasizes the reuse and re-engineering of existing ontologies, fostering collaboration between ontology engineers and domain experts to facilitate a more comprehensive and efficient development process. UPON is a methodology that incorporates concepts from the Unified Modeling Language (UML) and the Rational Unified Process (RUP). This approach adheres to an iterative and incremental process, encompassing four development phases: inception, elaboration, construction, and transition. It provides a structured framework for creating and refining ontologies, ensuring consistent and systematic ontology development. SAMOD is a methodology that integrates agile software development principles into ontology engineering. It provides a flexible, iterative, and adaptable framework, fostering collaboration among ontology engineers, domain experts, and end-users. Employing iterative development cycles, SAMOD encompasses planning, design, implementation, and evaluation stages, facilitating ongoing feedback and refinements throughout the development process. Finally, MoMo is a methodology that creates and manages modular ontologies, allowing for independent development, maintenance, and reuse of smaller, more manageable modules. It facilitates collaboration and focuses on specific parts of the ontology. MoMo involves module identification, design, integration, and evaluation, resulting in a more efficient and maintainable ontology development process. In general, the use of an ontology development methodology can lead to more efficient ontology development, better ontology reuse, easier ontology maintenance, and more effective collaboration between ontology engineers and domain experts. Furthermore, the adoption of a standardized ontology development methodology can improve the interoperability and compatibility of ontologies, enabling more effective data integration and knowledge sharing among different systems and applications.

The selection of an appropriate ontology development methodology depends on several factors, such as the complexity of the ontology, the domain of application, the availability of resources, and the expertise of the ontology engineers. In this work, we have opted for SAMOD to build a user contextual profile ontology due to its flexibility, adaptability, and emphasis on collaboration among ontology engineers, domain experts, and end-users. The iterative and incremental process of SAMOD, combined with its ongoing feedback and refinements, aligns with our goal of creating a comprehensive and adaptable ontology that can evolve over time. Hence, we will describe the development process of our user contextual profile ontology in the next section.
\section{Methodology for developing a user contextual profile ontology}\label{sec03ourapproach}
The development of a User Context Profile Ontology is a complex process that requires a structured and systematic approach. SAMOD methodology is based on agile software development principles, enabling the efficient development of a comprehensive and adaptable UCPO. Therefore, the UCPO can be made dynamic through continuous updates and modifications to the ontology based on changes in user profiles and contextual information. This can be achieved by incorporating an agile approach to ontology development, where the ontology is built and adapted incrementally, with regular feedback and updates from users. In this section, we describe the phases of SAMOD, including kickoff, design and implementation, and test and evaluation, which serve as a framework for designing the UCPO.
\subsection{Kickoff phase}
The main objective of the UCPO is to establish a standardized representation of user information and contextual information for each user profile. This ontological user modeling can facilitate the development of personalized systems and services, including personalized information retrieval, adaptive user interfaces, personalized recommendations, and other applications. Defining a standardized format for user information, context, and profile can improve the interoperability of the UCPO and enable it to adapt to the changing needs and preferences of the users.
The scope of the UCPO is broad, covering a wide range of user characteristics and environmental factors that can influence user behavior and preferences. This includes, but is not limited to, demographics, psychographics, device and network context, physical context, social context, and temporal context. The ontology will be designed to be flexible and adaptable, allowing for the incorporation of new or changing user characteristics and environmental factors.

To develop the UCPO, we consult a variety of information sources and references such as academic articles, conference proceedings, textbooks, and online resources related to ontology engineering, user modeling, and personalization. In selecting the information sources, it is important to consider their relevance to the scope and goals of the UCPO, as well as the quality and credibility of the sources. We manually reviewed and analyzed the relevant literature to identify the key concepts and terms related to user context and personalization. Additionally, consulting domain experts, ontology engineers, and end-users can provide valuable insights into the specific user characteristics and contextual factors that should be incorporated into the ontology.
\begin{table}
	\begin{center}
	\caption{Examples of informal competency questions
		tables.}\label{tab_01}
\begin{tabular}{|P{1cm}|l|}
	\hline
	\textbf{ID} &  \quad\textbf{Questions}\\
	\hline
	CQ1 &  \:\:What is demographic information of the user ?\\\hline
	CQ2 &  \:\:What is the user's preferred vehicle type?\\\hline
	CQ3 &  \:\:What is the user's budget for a vehicle purchase?\\\hline
	CQ4 &  \:\:Which particular vehicle models are favored by the user?\\\hline
	CQ5 &  \:\:What is the user's driving environment?\\\hline
	CQ6 &  \:\:What is the user's preferred vehicle brand?\\\hline
	CQ7 &  \:\:What are the primary use cases for a particular vehicle model?\:\:\\\hline
	CQ8 &  \:\:What is the user's preferred vehicle transmission type \\\hline
\end{tabular}
\end{center}
\end{table}

The ontology requirements can be formulated by using informal competency questions. The set of competency questions helps to identify and define the key concepts, relationships, and attributes that should be included in the ontology. Competency questions serve as a guide for ontology development and help to ensure that the resulting ontology accurately represents the user context and profile information required for personalization. As in the table \ref{tab_01}, we illustrate some informal competency questions related to  define the key characteristics of the user contextual profile in the vehicle sales domain, which can then be used to inform the development of the UCPO. 

Defining the goal, scope, information sources, and competency questions helps to establish a shared understanding of the primary requirements of the UCPO between domain experts and ontology engineers. In the next section, we will focus on the main phases of the development process, where the key components of the UCPO are designed and implemented to obtain a comprehensive and adaptable version.
\subsection{Design and implementation phase}
Considered as the core of the ontology engineering process, the design and implementation phase consists of several iterations of ontology design, implementation, and testing until a comprehensive and adaptable ontology is achieved. The design phase begins by identifying the key concepts and relationships that should be included in the ontology based on the competency questions formulated in the previous phase. SAMOD prescribes an iterative process that aims to build the final model through a series of small steps. One key component of this process is the use of concept modelets, which are standalone models that describe particular aspects of the domain under consideration. These modelets are used to provide an initial conceptualization without being constrained by the current model available after the previous iteration of the process.

\begin{figure}[h!]
	\includegraphics[width=\textwidth]{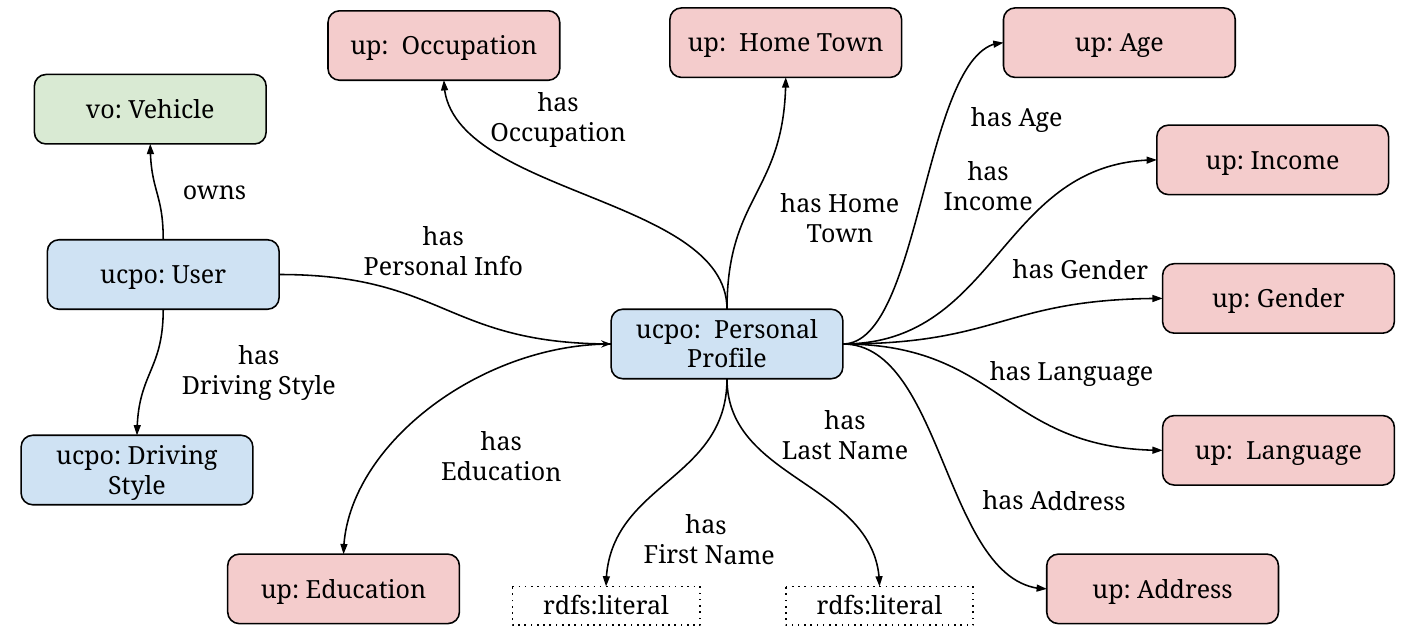}
	\caption{A dedicated ontology section for demographic and social information of users, composed of the User Profile Ontology (highlighted in red boxes with $up$ prefix), User Contextual Profile Ontology (highlighted in blue boxes with $ucpo$ prefix), and the vehicle domain ontology (highlighted in green boxes with $vo$ prefix).} \label{fig_01}
\end{figure}

We have organized the development of the UCPO into two separate modelets. The first modelet focuses on representing mostly static and permanent user characteristics. The second modelet presents both static and dynamic information about the user profile and contextual factors that affect the user and their profile. As shown in Figure \ref{fig_01}, the first modelet is responsible for organizing user demographic and social information, such as $Gender$, $Age$, $address$, $occupation$, $Income$, $Language$, $NumberOfChildren$, $MaritalStatus$, $Education$, and more. \textit{Gender} refers to a person's biological identity as male or female. \textit{Age} refers to the user's age, typically measured in years. \textit{Address} refers to a person's physical location, which can be used to gather contextual information about the user's environment. \textit{Occupation} refers to a person's job or profession, which can provide insights into their daily routine and transportation needs. \textit{Income} refers to a person's financial status, which can affect their purchasing power and preferences. \textit{Language} refers to a person's preferred language for communication, which can influence the type of content and recommendations that they receive. \textit{NumberOfChildren} refers to the number of children a person has, which can impact their vehicle preferences and needs. \textit{MaritalStatus} refers to a person's current marital status, which can provide insights into their family structure and needs. \textit{Education} level refers to the highest level of education a person has completed, which can be a factor in determining their interests and preferences.  In general, these types of information have been well-organized into classes in many previous works, allowing for the description of user information in a categorical manner. In our research, we utilized the work of Bermudez et al. \cite{upo} for the basic classes of user information. Moreover, we have created a class named $PersonalProfile$, which acts as a superclass for the classes related to the demographic and social information. The $PersonalProfile$ class can contains attributes such as first name, last name of the user.

\begin{figure}[h!]
	\includegraphics[width=\textwidth]{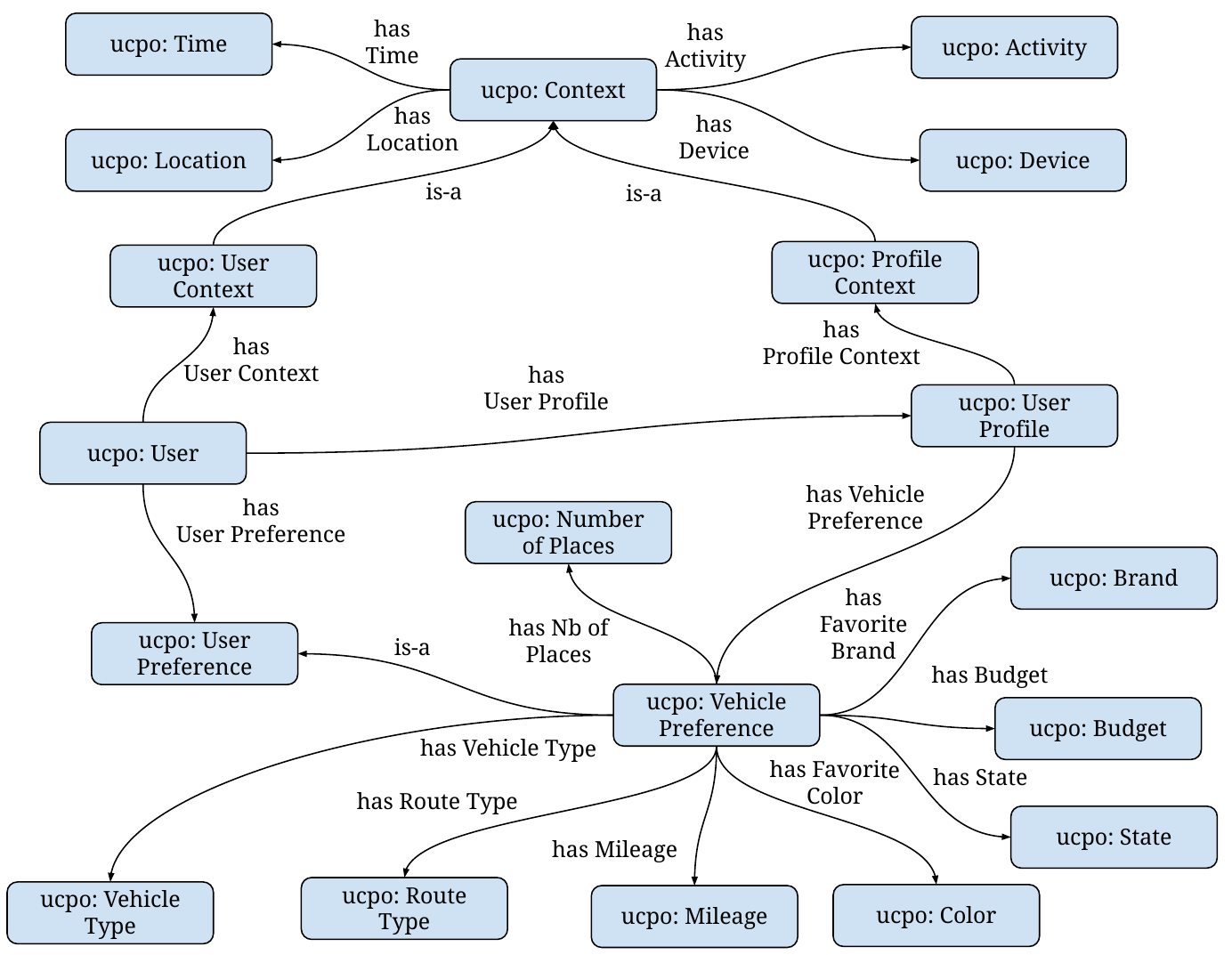}
	\caption{A dedicated ontology section for user preferences and contextual information of user and each particular profile.} \label{fig_02}
\end{figure}

The second modelet focuses on the representation of the structure and relationships between a user, their user profile, and their user context. User context refers to the circumstances, environment, and situation in which a user interacts with a system or application. It includes a wide range of factors such as the user's location, time, device, preferences, goals, and social context, among others. User context is crucial because it can have a significant impact on the user's behavior, preferences, and needs. As shown in Figure \ref{fig_02}, we investigated four particular contextual information about the user, including $Time$, $Location$, $Activity$, and $Device$ classes. In our design, we created two subclasses of $Context$ class: the $UserContext$ class that describes general contextual information about the user and the $ProfileContext$ class, which describes contextual information for each user profile. Furthermore, we have provided the ability for each user to declare their preferences and create various types of profiles that align with their intended purpose while interacting with the system or application. To structure this type of information, we have created two classes: $Preference$ and $UserProfile$.

In order to address the needs and preferences of users in the vehicle sales domain, we have created a set of preferences that relate to the user's desired vehicle. This set of preferences includes several classes such as $VehicleType$, $RouteType$, $Mileage$, $Color$, $NumberOfPlaces$, $State$, $Budget$, and $Brand$. The $VehicleType$ class refers to the type of vehicle the user is interested in, such as sedan, SUV, or truck. The $RouteType$ class describes the user's preferred driving route, such as highway or city streets. The $Mileage$ class indicates the user's desired mileage range. The $Color$ class refers to the user's preferred color for their vehicle. The $NumberOfPlaces$ class indicates the number of seats or passengers the user wants in their vehicle. The $State$ class represents the vehicle state where the user intends to purchase. The $Budget$ class refers to the user's budget for the vehicle purchase. The $Brand$ class represents the user's preferred vehicle brand.

It is important to note that depending on the specific domain and application, additional preferences may be necessary to fully capture the user's needs and preferences. Therefore, sub-classes of $Preference$ class can be created and customized based on the specific requirements of the vehicle sales domain and the user base. By incorporating these preferences into the user contextual profile ontology, we can provide a more personalized and tailored user experience.

Using the conceptual model, we can outline the major classes, attributes, and relationships between them (as illustrated in Figures \ref{fig_01} and \ref{fig_02}). This model is then refined through multiple iterations, taking feedback from stakeholders and end-users into consideration to ensure that it accurately represents the domain knowledge and requirements.

\begin{figure}[h!]
	\centering\includegraphics[width=\textwidth]{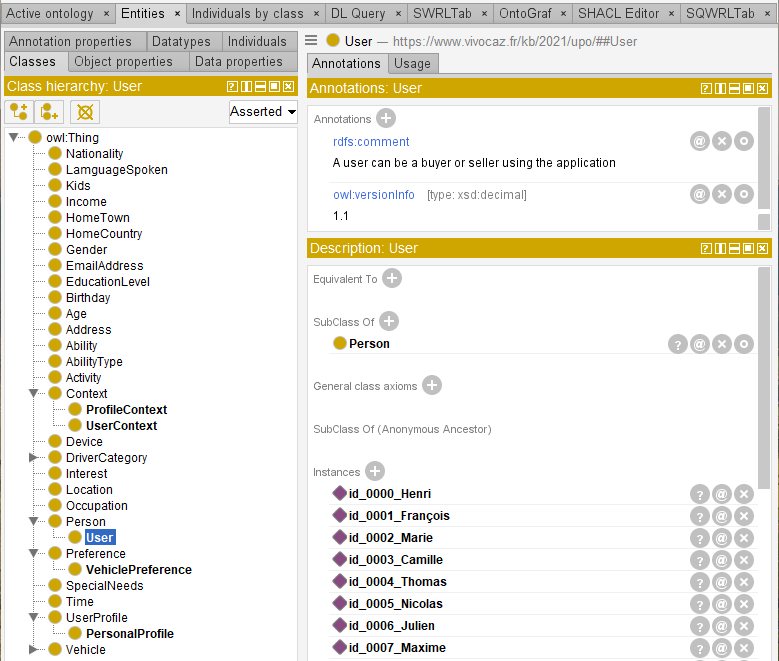}
	\caption{A snapshot of the User Contextual Profile Ontology in the Protégé-OWL editor} \label{fig_03}
\end{figure}

The implementation of the UCPO involves the process of actually creating and deploying the ontology in a system or application. This requires converting the conceptual model of the ontology into a computer-readable format using ontology languages such as OWL or RDF. As shown in Figure \ref{fig_03}, we have implemented the UCPO using the Web Ontology Language (OWL) with the support of the Protégé-OWL editor \cite{musen2015protege}.

After the design and implementation phase, the ontology undergoes further refinement through multiple rounds of testing and feedback until a final version is achieved. The ontology is then evaluated to ensure it meets the intended goals and requirements and can be integrated with existing systems and applications. In the next section, we present the test and evaluation phase of the ontology.



\subsection{Test and evaluation phase}
In the final phase of the UCPO development, we focus on testing and evaluating the produced ontology. The tests can be categorized into a model test, a data test, and query tests. To perform the model test, we can employ evaluation metrics such as OOPS! \cite{poveda2014oops}, OntoQA \cite{tartir2010ontological}, and OntoMetrics \cite{10.5220/0006084601860191} to verify the overall consistency of the developed ontology. The data test consists of checking the validity of the model after populating it with instance triplets. Finally, for query tests, informal competency questions must be transformed into SPARQL queries to ensure that the expected answers are obtained. The ontology model must be adjusted until all tests are successful. This section presents our method for conducting these types of tests to ensure the ontology's adaptation. 

In our work, we evaluated the quality of our ontology by examining its structure. Specifically, we observed several metrics using the OntoMetrics framework. Table \ref{tab:tb_3} presents the results from the analysis of OntoMetrics with base metrics and schema metrics. The base metrics consist of simple metrics that count the number of axioms, classes, object properties, data properties, and individual instances. These metrics provide statistics about the ontology with regard to the quantity of elements present within it. Based on these statistics, we can conclude that our ontology is a lightweight knowledge graph that can be adopted in different application architectures. Moreover, the Description Logic (DL) expressivity of the ontology is $\mathcal{ALH}(D)$, which indicates the DL variant that the ontology belongs to. $\mathcal{ALH}(D)$ refers to attribute language $\mathcal{AL}$ with role hierarchy ($\mathcal{H}$).

\begin{table}[h]
	\centering
	
	\caption{Results achieved from applying various metrics to UCPO.}
	\label{tab:tb_3}
	\begin{tabular}{|l|c|}
		\hline
		\multicolumn{2}{|c|}{ \textbf{Base Metrics}}\\\hline
		\:\:Class count & 38 \\\hline
		\:\:Object property count & 27\\\hline
		\:\:Data property count & 16 \\\hline
		\:\:Properties count & 43\\\hline
		\:\:Individual count & 159\\\hline
		\:\:SubClassOf axioms count& 22\\\hline
		\:\:Object property domain axioms count\:\: & 28\\\hline
		\:\:Object property range axioms count & 33 \\\hline
		\:\:DL expressivity & 	$\mathcal{ALH}(D)$ \\\hline
		\multicolumn{2}{|c|}{ \textbf{Schema Metrics}}\\\hline
		\:\:Attribute richness (AR) & 0.421053 \\\hline
		\:\:Inheritance richness (IR) & 0.578947 \\\hline
		\:\:Relationship richness (RR) & 0.55102 \\\hline
		\:\:Axiom/class ratio & \:\:30.552632\:\:\\\hline
		\:\:Class/relation ration & 0.77551 \\\hline
	\end{tabular}
\end{table}

In schema metrics,  the evaluation of the ontology focuses on the measure of attribute, inheritance, and relationship richness \cite{tartir2010ontological}. Particularly, the attribute richness ($AR$) is considered as the average number of attributes per class and is computed as follows: 
\begin{equation}
	AR = \frac{|NA|}{|C|}
\end{equation} where  $NA$ is the  number attributes for all classes and $C$ denotes the number  of classes. The high score of the attribute richness indicates the high quality of ontology design and the amount of information belong to instances. The inheritance richness ($IR$) is defined as the average number of sub-classes per class and is computed as follows:
\begin{equation}
	IR = \frac{|H|}{|C|}
\end{equation} where $H$ is the sum of the number of inheritance relationships. The inheritance richness allows describing the distribution of information on different levels of the ontology. The relationship richness ($RR$) is defined as the percentage of the relationships between classes and is computed as follows:
\begin{equation}
	RR = \frac{|P|}{|H| + |P|}
\end{equation} where $P$ the sum of the number of non-inheritance relationships. The inheritance richness is used to measure the diversity of the relationship's types in the ontology. 
Moreover, the other metrics such as  axiom/class ratio and class/relation ratio demonstrate the ratio between axioms and classes,  and classes-relations. Based on the explanation of metrics, these measured results in table \ref{tab:tb_3} indicate that our proposed ontology is balanced between a horizontal (or shallow) ontology  and vertical (or deep) ontology.

To conduct a data test for the UCPO, we need to check the validity of the model after populating it with instance triplets. For example, let's assume we have created instances for Henri's two profiles: his professional profile and his family profile. We can populate these instances with data such as his occupation and work location for his professional profile, and his number of children and preferred family activities for his family profile. This ensures that the UCPO can capture and express this information in a structured and consistent manner. 

In order to conduct a query test on the UCPO, we can convert the set of informal competency questions, as shown in Table \ref{tab_01}, into SPARQL queries to ensure that the expected answers are obtained. The queries should be constructed in a way that enables the retrieval of relevant information from the ontology, utilizing the appropriate classes, properties, and instances. For instance, if the competency question is ``What is the user's preferred vehicle brand?'', a suitable SPARQL query can be formulated as follows:
\begin{lstlisting}[captionpos=b, caption=A SPARQL query expression used to search for the first 10 users and their favorite brands., label=lst:sparql,
	basicstyle=\scriptsize\ttfamily,frame=lines]
PREFIX vo: <http:// vivocaz.fr/vo/ns#>
PREFIX ucpo: <http:// vivocaz.fr/ucpo/ns#>
SELECT ?user ?brand
WHERE {
	?user ucpo:hasUserProfile ?userProfile.
	?userProfile ucpo:hasVehiclePreference ?userVehiclePreference . 
	?userVehiclePreference upo:hasFavoriteBrand ?brand .
} ORDER BY ?user LIMIT 10
\end{lstlisting}

The test and evaluation phase ensures that the UCPO is consistent, valid, and effective in achieving its intended purpose. By conducting model tests, data tests, and query tests, we can evaluate the overall structure of the ontology, the validity of its populated instances, and its ability to accurately answer competency questions. In the next section, we will explore an application of the UCPO through a case study focused on generating personalized recommendations.

\section{Case study - Personalized Recommendations}\label{sec04experiments}
To demonstrate the application of the UCPO, we conducted a case study on personalized recommendations in the vehicle sales domain. Using the ontology, we were able to capture the user's contextual information, preferences, and profiles to provide tailored recommendations for vehicle purchases. Specifically, scenarios were considered where a user created a profile specifying their preferred $vehicle$ $type$, $route$ $type$, $mileage$, $color$, $number$ $o$f $seats$, $location$, $brand$, and $budget$. Additionally, the user's profile context included their $activity$, $time$ $of$ $day$, and $device$. This information was used to provide personalized recommendations for vehicle purchases that catered to the user's individual requirements and preferences.

Louis and Pierre are two potential buyers of used vehicles who have registered on a vehicle sales application and set up their profiles. Louis has specified his preferences for $vehicle$ $type$, $route$ $type$, $mileage$, $color$, $number$ $of$ $seats$, $location$, $brand$, and $budget$, while expressing a particular interest in sedan models. On the other hand, Pierre has emphasized the importance of safety features as a top priority due to frequent travel with his family.

The application integrates the UCPO to capture their contextual information and preferences. By leveraging Louis's contextual information and preferences, including his activity or location such as the Peugeot models he has previously liked, the application generates personalized recommendations for vehicle purchases that cater to his specific needs and preferences. The application recommends models such as $Peugeot$ $206$, $Peugeot$ $207$, or $Peugeot$ $208$ that Louis may like, based on his stated preferences and his profile context.

On the other hand, the application captures Pierre's contextual information and preferences, including his activity or location such as the fact that he previously owned a $Toyota$ and expressed interest in hybrid vehicles, to provide tailored recommendations. By leveraging this information, the application's recommendation engine suggests several vehicles that fit Pierre's criteria, including a $Toyota$ $RAV4$ $Hybrid$, a $Honda$ $CR-V$ $Hybrid$, and a $Lexus$ $UX$ $Hybrid$. The system takes into account factors such as safety ratings, fuel efficiency, and cost to make the best relevant recommendations for Pierre.

Consider the example of Henri, who was mentioned in the introduction section, and has two distinct profiles: one as a professional profile and the other as a family profile. It is essential to provide personalized recommendations for Henri that meet his specific needs and preferences in each profile. By leveraging the UCPO, which captures user profiles and contextual information of each profile, relevant recommendations can be provided that cater to each corresponding profile. Specifically, for Henri's professional profile who has shown interest in Renault models, vehicles such as $Renault$ $Megane$ and $Renault$ $Talisman$, which have spacious trunks, advanced safety features, and good fuel efficiency, could be recommended, as Henri may require frequent business trips and the recommendations match the contextual information of his activities on the application. For Henri's family profile, who has indicated interest in SUV vehicle types, vehicles such as $Koleos$ $SUV$ and $Renault$ $Scenic$, which have ample space and storage and rear-seat airbags for passengers, could be recommended as Henri has shown a preference for these models in the family profile on the applications.

These scenarios illustrate how the UCPO ontology can be utilized to capture contextual information and preferences to generate personalized recommendations that cater to the specific needs and preferences of each user. By analyzing and understanding user information, systems or applications can provide more accurate and effective recommendations, as demonstrated in various works \cite{le2023constraint,luyen2023towards}.
By leveraging the UCPO in this way, it is fully potential to develop a system or application that provides more personalized recommendations tailored to the user's unique needs, preferences, and contextual information.

\section{Conclusion and Perspectives}\label{sec05conclusion}

In this paper, we have presented the design and development process of the User Contextual Profile Ontology with a focus on the vehicle sales domain. This work established the bridge for the gap in the representation of user profile and user context by designing a user contextual profile ontology that considers both user profiles and contextual information on each profile. Thereby, the UCPO serves as a structural framework for standardizing the representation of user profiles and contextual information. It is worth emphasizing that the development of the UCPO follows the main phases of ontology engineering, ensuring its consistency, validity, and effectiveness in achieving its intended purpose. The iterative development process of the UCPO involved testing and evaluation phases, which allowed for assessing the ontology's overall structure, validity of populated instances, and ability to answer competency questions accurately.  The case study presented illustrates the use of the UCPO in generating personalized recommendations in the vehicle sales domain. By leveraging the ontology's ability to capture user preferences and requirements accurately, the system can provide tailored recommendations that cater to the specific needs, preferences, and context of each user. In future work, we plan to extend the ontology to include additional information sources, such as user reviews and social media feeds, to enhance the accuracy of personalized recommendations. Finally, we aim to collaborate with industry partners to integrate the UCPO into commercial applications and evaluate its effectiveness in real-world settings.

\section*{Acknowledgment}
\vspace{-0.25cm}
This work was funded by the French Research Agency (ANR) and by the company Vivocaz under the project France Relance - preservation of R\&D employment (ANR-21-PRRD-0072-01).
\vspace{-0.25cm}
%
%
\bibliographystyle{splncs04}
\bibliography{references}

\begin{thebibliography}{10}
\providecommand{\url}[1]{\texttt{#1}}
\providecommand{\urlprefix}{URL }
\providecommand{\doi}[1]{https://doi.org/#1}

\bibitem{adewoyin2022user}
Adewoyin, O., Wesson, J., Vogts, D.: User modelling to support behavioural
  modelling in smart environments. In: 2022 3rd International Conference on
  Next Generation Computing Applications (NextComp). pp.~1--6. IEEE (2022)

\bibitem{de2005proposal}
De~Nicola, A., Missikoff, M., Navigli, R.: A proposal for a unified process for
  ontology building: Upon. In: Database and Expert Systems Applications: 16th
  International Conference, DEXA 2005, Copenhagen, Denmark, August 22-26, 2005.
  Proceedings 16. pp. 655--664. Springer (2005)

\bibitem{fernandez1997methontology}
Fern{\'a}ndez-L{\'o}pez, M., G{\'o}mez-P{\'e}rez, A., Juristo, N.:
  Methontology: From ontological art towards ontological engineering. In:
  Proceedings of the Ontological Engineering AAAI-97 Spring Symposium Series.
  American Asociation for Artificial Intelligence (1997)

\bibitem{he2008ontological}
He, S., Fang, M.: Ontological user profiling on personalized recommendation in
  e-commerce. In: 2008 IEEE International Conference on e-Business Engineering.
  pp. 585--589. IEEE (2008)

\bibitem{kourtiche2020oupip}
Kourtiche, A., mohamed Benslimane, S., Hacene, S.B.: Oupip: Ontology based user
  profile for impairment person in dynamic situation aware social networks.
  International Journal of Knowledge-Based Organizations (IJKBO)
  \textbf{10}(2),  12--34 (2020)

\bibitem{kritikou2008user}
Kritikou, Y., Demestichas, P., Adamopoulou, E., Demestichas, K., Theologou, M.,
  Paradia, M.: User profile modeling in the context of web-based learning
  management systems. Journal of Network and Computer Applications
  \textbf{31}(4),  603--627 (2008)

\bibitem{10.5220/0006084601860191}
Lantow, B.: Ontometrics: Putting metrics into use for ontology evaluation. In:
  Proceedings of the International Joint Conference on Knowledge Discovery,
  Knowledge Engineering and Knowledge Management. p. 186–191. IC3K 2016,
  SCITEPRESS - Science and Technology Publications, Lda, Setubal, PRT (2016),
  \url{https://doi.org/10.5220/0006084601860191}

\bibitem{le2022apport}
Le, N.L., Abel, M.H., Gouspillou, P.: Apport des ontologies pour le calcul de
  la similarit{\'e} s{\'e}mantique au sein d'un syst{\`e}me de recommandation.
  In: 33{\`e}me Journ{\'e}es Francophones d'Ing{\'e}nierie des Connaissances
  (IC 2022). pp. 189--198 (2022)

\bibitem{luyen2023towards}
Le, N.L., Abel, M.H., Gouspillou, P.: Towards an ontology-based recommender
  system for the vehicle sales area. In: Progresses in Artificial
  Intelligence {\&} Robotics: Algorithms {\&} Applications. pp. 126--136.
  Springer International Publishing, Cham (2022)

\bibitem{le2023personalized}
Le, N.L., Abel, M.H., Gouspillou, P.: A personalized recommender system
  based-on knowledge graph embeddings. In: The 3rd International Conference on
  Artificial Intelligence and Computer Vision (AICV2023), March 5--7, 2023. pp.
  368--378. Springer (2023)

\bibitem{luyen2023}
Le, N.L., Abel, M.H., Gouspillou, P.: Improving semantic similarity measure
  within a recommender system based-on rdf graphs. In: Proceedings of the 6th
  International Conference on Information Technology \& Systems (2023)

\bibitem{luyen2016development}
Le, N.L., Tireau, A., Venkatesan, A., Neveu, P., Larmande, P.: Development of a
  knowledge system for big data: Case study to plant phenotyping data. In:
  Proceedings of the 6th International Conference on Web Intelligence, Mining
  and Semantics. pp.~1--9 (2016)

\bibitem{le2023constraint}
Le, N.L., Zhong, J., Negre, E., Abel, M.H.: Constraint-based recommender system
  for crisis management simulations. In: The 56th Hawaii International
  Conference on System Sciences (2023)

\bibitem{li2019using}
Li, S., Abel, M.H., Negre, E.: Using user contextual profile for recommendation
  in collaborations. In: Research \& Innovation Forum 2019: Technology,
  Innovation, Education, and their Social Impact 1. pp. 199--209. Springer
  (2019)

\bibitem{upo}
Maria, B., Payam, B., Sefki, K.: User profile ontology (2015),
  \url{http://iot.ee.surrey.ac.uk/citypulse/ontologies/up}

\bibitem{maria2007creating}
Maria, G., Akrivi, K., Costas, V., George, L., Constantin, H.: Creating an
  ontology for the user profile: Method and applications. In: Proceedings AI*
  AI Workshop RCIS. pp. 407--412 (2007)

\bibitem{munoz2021gente}
Mu{\~n}oz, H.J.M., Cardinale, Y.: Gente: An ontology to represent users in the
  tourism context. In: 2021 XLVII Latin American Computing Conference (CLEI).
  pp. 1--10. IEEE (2021)

\bibitem{musen2015protege}
Musen, M.A.: The prot{\'e}g{\'e} project: a look back and a look forward. AI
  matters  \textbf{1}(4),  4--12 (2015)

\bibitem{noy2001ontology}
Noy, N.F., McGuinness, D.L., et~al.: Ontology development 101: A guide to
  creating your first ontology (2001)

\bibitem{peroni2017simplified}
Peroni, S.: A simplified agile methodology for ontology development. In: OWL:
  Experiences and Directions--Reasoner Evaluation: 13th International Workshop,
  OWLED 2016, and 5th International Workshop, ORE 2016, Bologna, Italy,
  November 20, 2016, Revised Selected Papers 13. pp. 55--69. Springer (2017)

\bibitem{poveda2014oops}
Poveda-Villal{\'o}n, M., G{\'o}mez-P{\'e}rez, A., Su{\'a}rez-Figueroa, M.C.:
  Oops!(ontology pitfall scanner!): An on-line tool for ontology evaluation.
  International Journal on Semantic Web and Information Systems (IJSWIS)
  \textbf{10}(2),  7--34 (2014)

\bibitem{rimitha2019improving}
Rimitha, S., Abburu, V., Kiranmai, A., Marimuthu, C., Chandrasekaran, K.:
  Improving job recommendation using ontological modeling and user profiles.
  In: 2019 Fifteenth International Conference on Information Processing
  (ICINPRO). pp.~1--8. IEEE (2019)

\bibitem{shimizu2021modular}
Shimizu, C., Hammar, K., Hitzler, P.: Modular ontology modeling. Semantic Web
  pp. 1--31 (2021)

\bibitem{sieg2007web}
Sieg, A., Mobasher, B., Burke, R.: Web search personalization with ontological
  user profiles. In: Proceedings of the sixteenth ACM conference on Conference
  on information and knowledge management. pp. 525--534 (2007)

\bibitem{skillen2012ontological}
Skillen, K.L., Chen, L., Nugent, C.D., Donnelly, M.P., Burns, W., Solheim, I.:
  Ontological user profile modeling for context-aware application
  personalization. In: Ubiquitous Computing and Ambient Intelligence: 6th
  International Conference, UCAmI 2012, Vitoria-Gasteiz, Spain, December 3-5,
  2012. Proceedings 6. pp. 261--268. Springer (2012)

\bibitem{sosnovsky2010ontological}
Sosnovsky, S., Dicheva, D.: Ontological technologies for user modelling.
  International Journal of Metadata, Semantics and Ontologies  \textbf{5}(1),
  32--71 (2010)

\bibitem{suarez2011neon}
Su{\'a}rez-Figueroa, M.C., G{\'o}mez-P{\'e}rez, A., Fern{\'a}ndez-L{\'o}pez,
  M.: The neon methodology for ontology engineering. In: Ontology engineering
  in a networked world, pp. 9--34. Springer (2011)

\bibitem{sutterer2008upos}
Sutterer, M., Droegehorn, O., David, K.: Upos: User profile ontology with
  situation-dependent preferences support. In: First International Conference
  on Advances in Computer-Human Interaction. pp. 230--235. IEEE (2008)

\bibitem{sutterer2008user}
Sutterer, M., Droegehorn, O., David, K.: User profile selection by means of
  ontology reasoning. In: 2008 Fourth Advanced International Conference on
  Telecommunications. pp. 299--304. IEEE (2008)

\bibitem{tamine2010evaluation}
Tamine-Lechani, L., Boughanem, M., Daoud, M.: Evaluation of contextual
  information retrieval effectiveness: overview of issues and research.
  Knowledge and Information Systems  \textbf{24},  1--34 (2010)

\bibitem{tartir2010ontological}
Tartir, S., Arpinar, I.B., Sheth, A.P.: Ontological evaluation and validation.
  In: Theory and applications of ontology: Computer applications, pp. 115--130.
  Springer (2010)

\bibitem{wang2004ontology}
Wang, X.H., Zhang, D.Q., Gu, T., Pung, H.K.: Ontology based context modeling
  and reasoning using owl. In: IEEE annual conference on pervasive computing
  and communications workshops, 2004. Proceedings of the second. pp. 18--22.
  Ieee (2004)

\end{thebibliography}
\end{document}